\documentclass{sig-alternate}

\usepackage[pdftex,colorlinks=false,urlcolor=blue,citecolor=black,anchorcolor=black,linkcolor=black]{hyperref}
 
\begin{document}

\title{Attack of the Ants: Studying Ant Routing Algorithms in \\Simulation and Wireless Testbeds}
\numberofauthors{2}
\author{
\alignauthor Michael Frey\\
\affaddr{Department of Computer Science}\\
\affaddr{Humboldt-Universit{\"a}t zu Berlin}\\
\affaddr{Unter den Linden 6, 10099 Berlin}\\
\email{frey@informatik.hu-berlin.de}
\alignauthor Mesut G{\"u}nes\\
\affaddr{Institute of Computer Science}\\
\affaddr{University of M{\"u}nster}\\
\affaddr{Einsteinstrasse 62, 48149 M{\"u}nster}\\
\email{mesut.guenes@uni-muenster.de}
}      

\date{\today}

\maketitle
\begin{abstract}
Wireless networks are becoming the key building block of our communications infrastructure.
Examples range from cellular networks to ad hoc and sensor networks in wildlife monitoring and environmental scenarios.
With the rise of the Internet of Things (IoT) millions of physical and virtual objects will communicate wireless and enhance the daily life.
The adaptivity and scalability of wireless networks in the IoT is one of the most challenging tasks.
Bio-inspired networking algorithms are a way to tackle these issues.
In this paper we present a simulation framework based on OMNeT++ to implement ant routing algorithms to study and compare them on the algorithmic level and an approach to run large simulation studies in a comprehensive way.
\end{abstract}

%A category including the fourth, optional field follows...
%\category{C.2.1}{Network Architecture and Design}{Wireless communication}
%\keywords{ACM proceedings, \LaTeX, text tagging} % NOT required for Proceedings

\section{Introduction}
Wireless networks are becoming a important part of our communication infrastructure.
There is a wide variety ranging from cellular networks in mobile communications to ad hoc networks between cars or in wireless sensor networks (WSNs).
In addition, with the rise of the Internet of Things (IoT)~\cite{atzori+:2010} there is a shift in paradigms to the interconnection of everydays physical objects.
Though, the IoT faces new problems and challenges.
This includes but is not limited to the management and maintenance of million devices or providing a scalable, and flexible communication.
%Particularly providing a scalable and flexible communication is of utmost importance.
In addition, devices in the IoT are typically constrained in their computational power, memory, and energy.
Since these devices share commonalities with devices in WSNs, it seems applicable to apply algorithms from the area of WSNs to the IoT.
What specific WSNs algorithms are suited for the IoT remains an open research question.
However, researchers~\cite{Liotta:2013} suggest to face the challenges of the IoT using bio-inspired algorithms.
Bio-inspired algorithms exhibit properties of self-organization~\cite{Dressler:2007}, such as that they rely only on local knowledge and interact locally without central control.  
One class of bio-inspired algorithms are swarm-intelligence algorithms.
Communication in the IoT will be in most cases based on wireless technology.
Studies have shown that the uncertainties of the environment have a severe impact on wireless communication.
Including and covering these uncertainties in models used in simulation is a complex task and comes at its price. 
We believe that it is feasible to study different aspects of swarm intelligence algorithms using different techniques.
In order to investigate the robustness of swarm intelligence algorithms in face of the uncertainties of the environment we rely on wireless testbeds.
However, aspects of scalability or mobility are best studied in simulation.
Hence, we want to provide a methodology and framework to investigate different aspects of swarm intelligence algorithms using different techniques.
We focus in our work on routing algorithms based on the ant-colony optimization (ACO) metaheuristic. 
The ACO metaheuristic is inspired by the foraging behavior of ants, where ants mark favorable paths towards a food source using a special hormone called pheromone. 
Since pheromones are exhibited to a natural evaporation process only the shortest path towards a food source remains.
The remainder of this paper is structured as follows. 
First, we discuss problems and challenges in studying ant routing algorithms for wireless networks in Section~\ref{sec:prob}.
We present our library on studying ant routing algorithms in Section~\ref{sec:libara}.
Subsequently, in Section~\ref{sec:methodology} we present our methodology to study ant routing algorithms in simulation and wireless testbeds. 
The paper concludes with a short summary.

\section{Problems and Challenges}\label{sec:prob}
Like other routing algorithms for wireless multi-hop networks (WMHNs), ant routing algorithms face the same problems and challenges if brought to the real world.
This includes but is not limited to uncertainties in the environment, or internal and external interferences.
Observed effects are unstable and asymmetric links which have a severe impact on the performance and overall functionality of ant routing algorithms.
In addition, researchers consider~\cite{Paquereau:2012} efficiency as the main challenge in ant routing for WMHNs.
Particularly, how an ant routing algorithm discovers and maintain routes and how fast it adapts to changes in a dynamic environment have a severe impact on the efficiency, both in terms of energy and usage of the wireless medium.
We also noticed that ant routing algorithms react sensitive to parameter settings. 
%Hence, a researcher has to carefully adjust the parameter settings in simulation.
%For example, Fig.~\ref{fig:phi} depicts the pheromone levels of two node disjoint paths with different weight settings in the transmission probability of an ant routing algorithm.
%\begin{figure}[ht]
%    \centering
%    \includegraphics[width=0.4\textwidth]{graphics/border-phi}
%    \caption{The architecture of \texttt{libARA} consisting of a core, simulation, and testbed component.}
%    \label{fig:architecture}
%\end{figure}
We believe that a methodology where we can study different aspects of ant routing algorithms will allow us to tackle the aforementioned issues. 

\section{A Library For Ant Routing Algorithms}\label{sec:libara}
The library for ant routing algorithms (\texttt{libARA})~\cite{frey+:2013b} enables researchers to study ant routing algorithms both in simulation and wireless testbeds. 
We implemented \texttt{libARA} in C++ and provide modular components for the design, implementation and evaluation of ant routing algorithms. 
The architecture of \texttt{libARA} has three main components: the core, the simulation and the testbed component.
%Fig.~\ref{fig:architecture} depicts a simplified version of the architecture of \texttt{libARA}. 
%\begin{figure}[ht]
%    \centering
%    \includegraphics[width=0.4\textwidth]{graphics/architecture}
%    \caption{The architecture of \texttt{libARA} consisting of a core, simulation, and testbed component.}
%    \label{fig:architecture}
%\end{figure}
The \texttt{libARA} core consists of three layers: concepts specific to ant routing algorithms, concepts independent to ant routing algorithms, and utilities. 
The concepts specific to ant routing algorithms include forwarding, evaporation, and reinforcement policies. 
The concepts independent of ant routing algorithms are abstractions for hardware access, addresses, and routing tables. 
Utilities provide methods and concepts supporting the study of ant routing algorithms. 
This includes but is not limited to logging and statistics mechanisms. 
We provide implementations of the \texttt{libARA} core for simulation and testbed. 
The simulation component uses OMNeT++ simulation core and the INETMANET package.
The testbed component uses libdessert, a underlay routing framework.
While there is no specific reason, one could provide an one implementation for simulation or testbed studies using only the core component of \texttt{libARA}. 
For a detailed description of \texttt{libARA} we refer the interesting reader to~\cite{frey+:2013b}.

\section{Methodology}\label{sec:methodology}
With the DES-Testbed we operate a wireless testbed for WMHNs with over 60 indoor and outdoor nodes. 
In addition, to the testbed infrastructure we operate also a virtualizer with the capacity to replicate all nodes of the testbed. 
On every virtual machine runs the same operating system and the same set of software that is available on the testbed. 
However, our software only emulates the wireless communication between the virtual machines.  
The focus of our methodology is to convey the insight from experiments in a particular environment to the next. 
\begin{figure}[ht]
    \centering
    \includegraphics[width=0.5\textwidth]{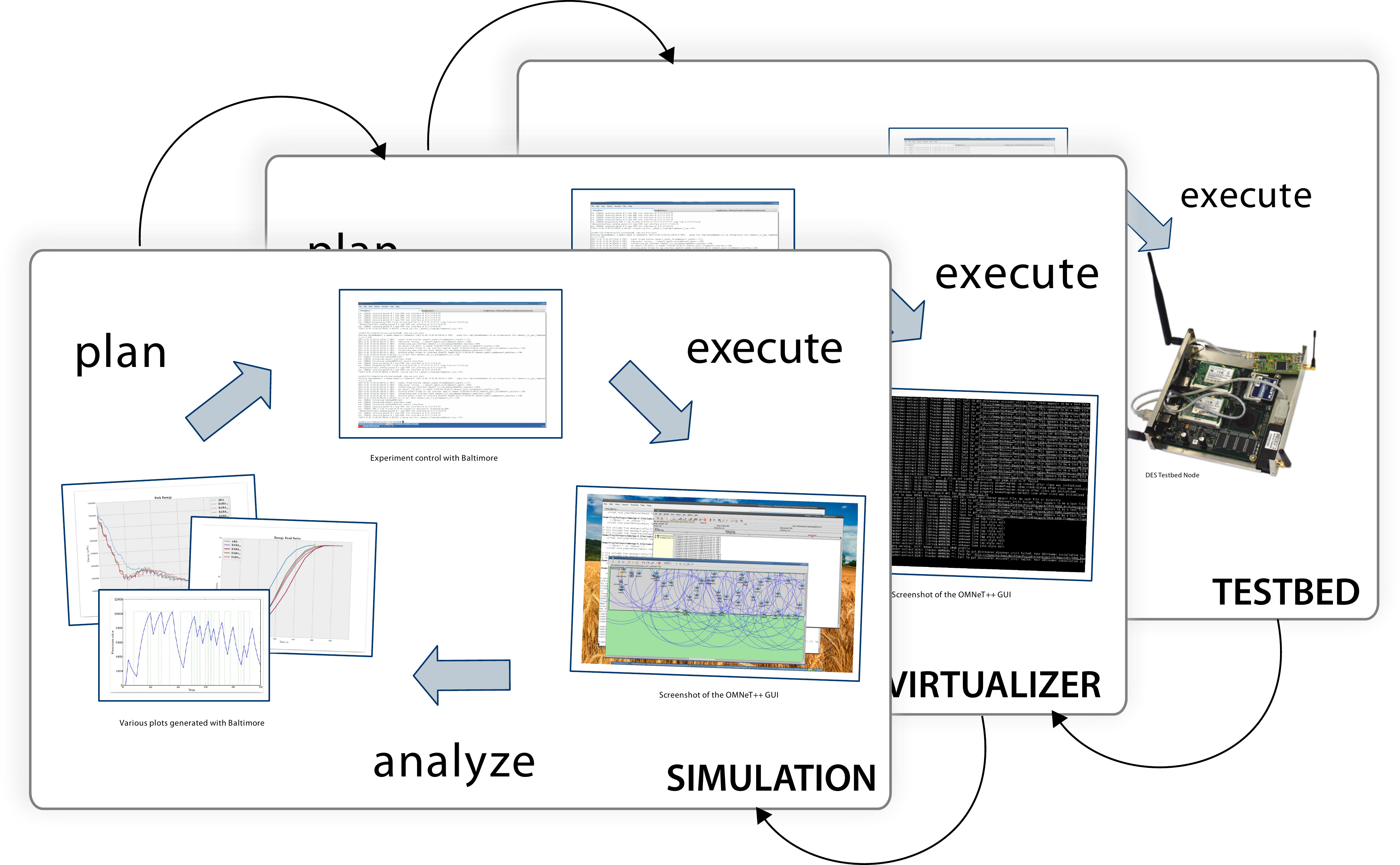}
    \caption{A scientific methodology to convey insights from one environment to the next.}
    \label{fig:method}
\end{figure}
Fig.~\ref{fig:method} shows our general methodology to study algorithms and protocols in large wireless multi-hop networks. 
We characterize an as a triple, e.g., environment = (hardware, software, radio transmission).
Each transition changes only one of these characteristics. 
From testbed to virtualizer the only change is the loss of real radio transmission. 
The transition from virtualizer to simulation consists of the loss of system software and the system environment.
Since each environment provides a particular advantage in the study of large multi-hop networks, this methodology supports the experimenter in a holistic study, i.e., realistic behavior of an algorithm in the testbed, dynamic behavior of an algorithm in the virtualizer, and scalability behavior in the simulation.

\section{Summary}
We presented in this paper our framework and methodology to study ant routing algorithms for wireless networks.
While running experiments in a wireless testbed is a cumbersome, expensive and error-prone task, studying ant routing algorithms in simulation allows to investigate certain properties and specifics of this algorithms more easily.
This includes behavioral aspects such as adaptivity and pheromone evolution, the scalability in respect to the number of nodes or traffic flows, and mobile scenarios. 
Our framework is easy to extend and to customize. 
Providing new backends for different network simulators (or testbed frameworks) is feasible with acceptable efforts. 
For the future we plan to provide additional ant routing algorithms. 
\texttt{libARA} is available from our github project page at \url{http://www.github.com/des-testbed/libara}.

\section*{Acknowledgements}
This work was partially supported with a research grant from the German Academic Exchange Service (Deutscher Akademischer Austausch Dienst, DAAD).

\bibliographystyle{abbrv}
\bibliography{references}

\balancecolumns
\end{document}